\documentclass[checkin,showpacs,aps,prl]{revtex4}
\usepackage{slashbox}

\newcommand{\bn}{\begin{enumerate}}
\newcommand{\en}{\end{enumerate}}
\newcommand{\ba}{\begin{eqnarray}}
\newcommand{\ea}{\end{eqnarray}}

\usepackage{graphicx,color}

\newcommand{\be}{\begin{equation}}
\newcommand{\ee}{\end{equation}}
\newcommand{\la}{\langle}

\newcommand{\ra}{\rangle}
\newcommand{\et}{{\it et al. }}
\newcommand{\ete}{{\it et al.}}

\def\prl{{ Phys. Rev. Lett. }}

\def\prb{{ Phys. Rev. B }}

\begin{document}

%\title{ Theory of ultrafast demagnetization:\\
%Perspectives from  spin-orbit-coupled Heisenberg system}

%\title{Demagnetization in a spin-orbit-coupled Heisenberg
%  system:\\ Applications to ultrafast spin dynamics}

%\title{Simple picture of demagnetization from a spin-orbit-coupled
%  Heisenberg system: From static to dynamic}

%\title{Essence of femtosecond demagnetization:\\ Exact results from a
%  spin-orbit-coupled Heisenberg system}

%\title{Understanding femtosecond magnetism:\\ A simple picture
%  from a spin-orbit-coupled Heisenberg system}

%\title{Exchange interaction  in femtomagnetism}
%\title{Electron exchange-interaction collapse as an alternative
%  mechanism for femtomagnetism }

%\title{A path to the consistent theory of femtosecond
%  magnetism:\\ Spin-orbit-coupled Heisenberg exchange model
%}

\title{Laser-induced ultrafast demagnetization time and spin moment in
  ferromagnets: First-principles calculation }

\author{G. P. Zhang$^*$} \affiliation{Department of Physics, Indiana
  State University, Terre Haute, IN 47809}

\author{M. S. Si} \affiliation{Key Laboratory for Magnetism and
  Magnetic Materials of the Ministry of Education, Lanzhou University,
  Lanzhou 730000, China}

 \author{Thomas F. George} \affiliation{
  Office of the Chancellor and Center for Nanoscience \\Departments of
  Chemistry \& Biochemistry and Physics \& Astronomy \\University of
  Missouri-St. Louis, St.  Louis, MO 63121 }

\date{\today}

\begin{abstract}
{ When a laser pulse excites a ferromagnet, its spin undergoes a
  dramatic change. The initial demagnetization process is very
  fast. Experimentally, it is found that the demagnetization time is
  related to the spin moment in the sample. In this study, we employ
  the first-principles method to directly simulate such a process. We
  use the fixed spin moment method to change the spin moment in
  ferromagnetic nickel, and then we employ the Liouville equation to
  couple the laser pulse to the system. We find that in general the
  dependence of demagnetization time on the spin moment is nonlinear:
  It decreases with the spin moment up to a point, after which an
  increase with the spin moment is observed, followed by a second
  decrease.  To understand this, we employ an extended Heisenberg
  model, which includes both the exchange interaction and spin-orbit
  coupling. The model directly links the demagnetization rate to the
  spin moment itself and demonstrates analytically that the spin
  relaxes more slowly with a small spin moment.  A future experimental
  test of our predictions is needed.  }
\end{abstract}


%ultrafast magnetization dynamics, 75.78.Jp

%75.40.Gb, 78.20.Ls, 75.70.-i, 78.47.J-}

%\keywords{femtomagnetism; exchange interaction}
 \maketitle

\newcommand{\tm}{\tau_m}
\newcommand{\ub}{\mu_{\rm B}}

More than a decade ago, Beaurepaire \et \cite{eric} demonstrated that
a femtosecond laser pulse could demagnetize a nickel thin film in less
than 1 picosecond, very attractive for future ultrafast magnetic
storage devices.  This inspired intensive experimental and theoretical
investigations
\cite{ourreview,rasingreview,prb98,prl00,lefkidis,fan,bigot,stanciu,lefkidismmm,carley,super,vah1,rhie},
with an even faster demagnetization reported \cite{chan,kra}. However,
what determines the demagnetization time $\tm$ is a complex and
challenging question, since multiple parameters, both intrinsic
(material properties) and extrinsic (laser pulse), jointly play a
role. With decades of experimental efforts, accurate demagnetization
times are now available at least for ferromagnetic $3d$ transition
metals. Bcc Fe has a spin moment of $M=2.2\ub$ and demagnetization
time of $\tm=98\pm 26$ fs \cite{mathias}, and fcc Ni has a spin moment
of $0.6\ub$ and $\tm=157\pm 9$ fs \cite{mathias} and 70-200 fs
\cite{koopmans}. Therefore, it appears that $\tm$ is inversely
proportional to $M$, but this trend is stopped at hcp Co. Its moment
is $1.6\ub$, but $\tm=160-240$ fs \cite{koopmans}, the longest among
the three. This highlights the challenge and complicated nature of
$\tm$, which is also affected by the exchange interaction.  Since the
demagnetization time is at the heart of the laser-induced ultrafast
demagnetization process, a detailed investigation is very appropriate
at this time.

In this paper, we focus on fcc Ni.  We employ the density functional
theory and the fixed spin moment method to gradually tune the
spin moment, without strongly affecting other parameters. Then we
carry out a dynamic simulation by solving the Liouville equation to
compute the dynamic change of the density matrices at each ${\bf k}$
point, from which the spin moment change is computed. We can determine
the demagnetization time for each spin moment. With an excitation by a
36 fs laser pulse, $\tm$ is found to decrease with the spin moment
until 0.35 $\ub$, after which a small increase is observed, before the
final decrease with the spin moment. The entire dependence appears
highly nonlinear, but the general trend is clear: $\tm$ decreases with
$M$. To explain this general trend, we adopt an extended Heisenberg
model, which includes both the exchange interaction and spin-orbit
coupling. We can analytically show that a larger spin moment indeed
leads to  a larger demagnetization rate, or shorter demagnetization
time.

\begin{figure}
\includegraphics[angle=270,width=14cm]{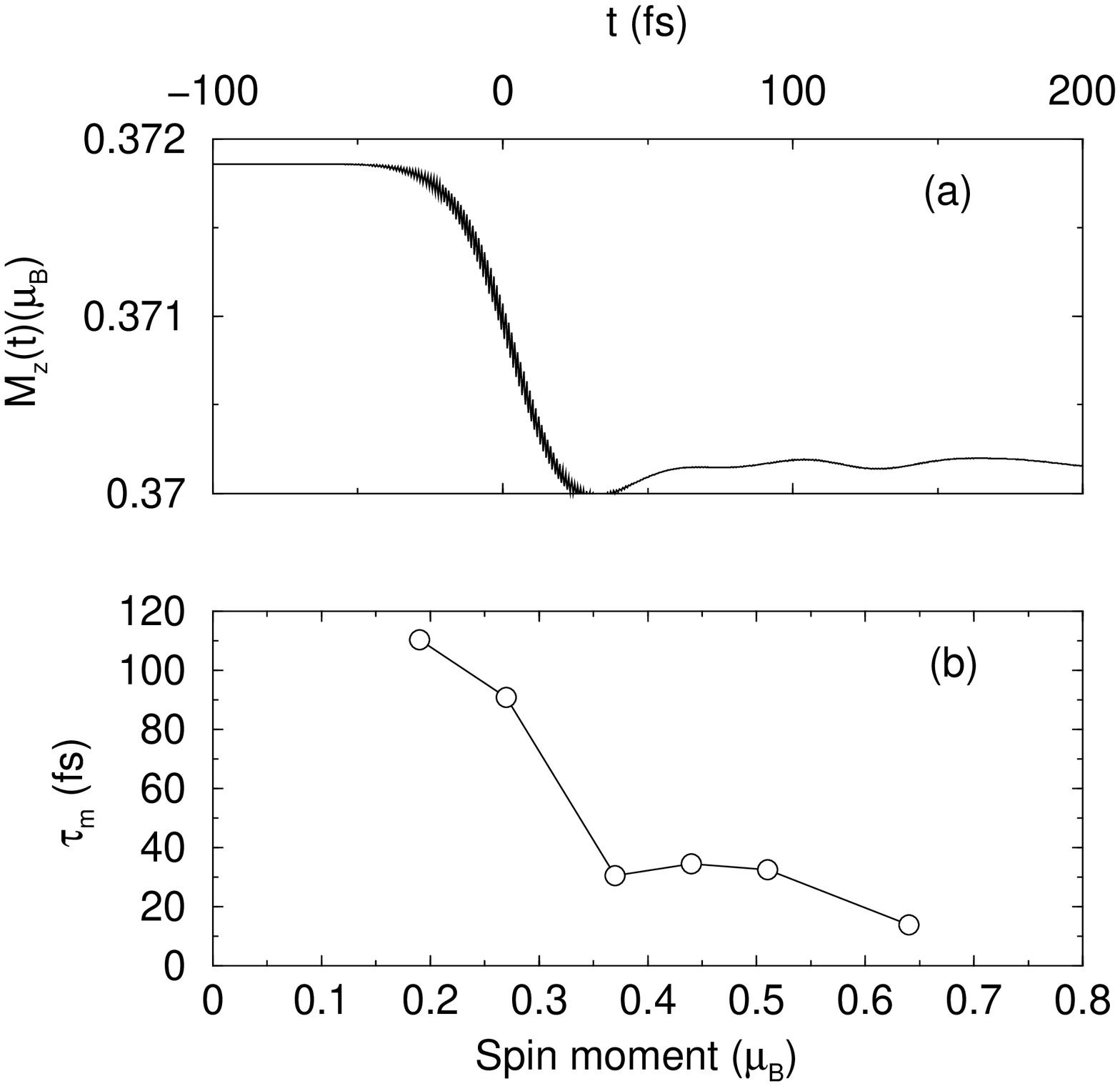}
\caption{ (a) Laser-induced ultrafast demagnetization, where the spin
  moment is 0.37 $\ub$. (b)
Demagnetization time $\tm$ as a function of spin moment. We choose six
different spin moments.
}
\label{fig1}
\end{figure}

To systematically investigate the effect of the spin moment on the
demagnetization time, we resort to the fixed spin moment method (FSM)
\cite{moruzzi}.  FSM was originally designed to study the complex
phase diagrams in complicated magnetic systems, where competing phases
coexist on a similar energy scale.  Within FSM, one starts from a
normal spin-polarized DFT calculation and computes the spin moment.
Then, one checks whether the spin moment matches the desired one. If
not, one refills the spin-up and spin-down states to get the desired
spin moment and forms a new charge density and potential for the next
iteration until convergence.  Different from the usual FSM, after the
self-consistency is reached, we perform an extra run including the
spin-orbit coupling, so we can investigate the spin moment change
under laser excitation.  We choose six spin moments, $0.19 \ub$, $0.27
\ub$, $0.37\ub$, $0.44 \ub$, $0.52 \ub$ and $0.64 \ub$. We use 104$^3$
${\bf k}$ points to converge our results.  Our real-time spin-moment
change is computed first by solving the Liouville equation
\cite{prb09,np09,jap08} \be i\hbar \left \la nk\left | \frac{\partial
  \rho}{\partial t} \right |mk\right \ra = \left \la nk\left |
     [H_0+H_I, \rho] \right |mk\right \ra, \ee where $|mk\ra$ is a
     band state $m$ at $k$. Once we obtain the density matrix, we
     trace over the product of the spin operator and density matrix to
     find the spin moment, $M_z={\rm Tr}(\rho (t) S_z)$.  Here $H_0$
     is the Hamiltonian for the system, and $H_I$ is the interaction
     Hamiltonian between the laser field and the system, \be
     H_I=\sum_{k}\sum_{nm} {\bf E ({\rm t})}\cdot {\bf
       D}_{k;nm}\rho_{k; nm}, \ee where ${\bf D}$ is the dipole
     operator. The laser field is \be {\bf E ({\rm t})} ={\bf A} {\rm
       e}^{-t^2/\tau^2} \cos(\omega t),\ee where we align the laser
     polarization ${\bf A}$ along the $x$-axis, and the magnetization
     quantization axis is along the $z$-axis. Our laser duration
     $\tau$ is 36 fs, the photon energy is $\hbar\omega=2$ eV, and the
     field amplitude $|{\bf A}|$ is 0.05 V/$\rm \AA$, which
     corresponds to a fluence of 11.5 mJ/cm$^2$, while experimentally
     the fluence ranges from 0.6 mJ/cm$^2$ \cite{kojpcm}
    to 35 mJ/cm$^2$ \cite{cheskis}. Figure \ref{fig1}(a) shows
     the spin moment changes as a function of time for the spin moment
     of 0.37 $\ub$. We notice that the spin drops very quickly upon
     the laser excitation, since the electrons in metals are mobile
     and can be easily excited out of the Fermi sea, as there is no
     gap blocking such a process, as expected from the Fermi liquid
     theory.  The time at the first spin minimum is defined as the
     demagnetization time $\tm$. For each spin moment, we carry out a
     similar calculation, and the detailed dependence on the spin
     moment is shown in Fig. \ref{fig1}(b). We find that the
     demagnetization time decreases very quickly with spin moment
     increase from 0.19 to 0.35 $\ub$. However, when the spin moment
     is above 0.35 $\ub$, it saturates, and interestingly, there is a
     small increase up to 0.45 $\ub$ before a further reduction with
     the spin moment. This dependence has never been reported before.

\begin{figure}
\includegraphics[angle=270,width=14cm]{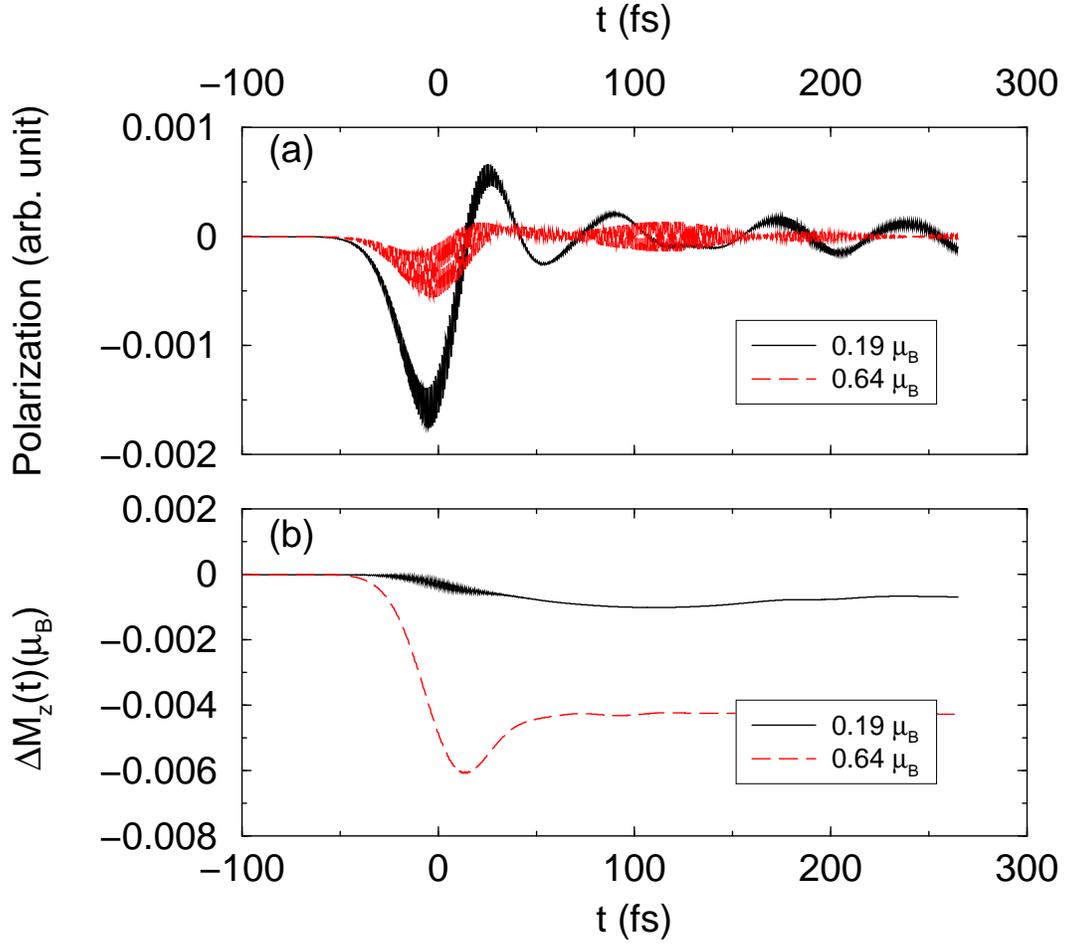}
\caption{
(a) Electric polarization change as a function of time for two spin
  moments at 0.19 (solid line) and 0.64$\ub$ (long dashed line).
(b) Spin moment change $\Delta M(t)$  as a function of time
  $t$. The notation of the lines is same as (a).
}
\label{fig2}
\end{figure}

We want to know whether the charge response already shows some
important differences. Figure \ref{fig2}(a) shows the electric
polarization as a function of time for the two spin moments of 0.19 (solid
line) and 0.64 (long-dashed line) $\ub$.  It is interesting that the
electric polarization behaves quite differently for these two
cases. Since in the present cases our electric and laser
polarizations are perpendicular to each other, the strong beating due
to the laser field is strongly suppressed, in comparison to the
collinear excitation \cite{jap08}. The excitation for 0.19 $\ub$ is
much stronger than 0.64 $\ub$ and has prominent oscillations. For 0.64
$\ub$, the oscillation is weaker. There is a clear decoupling between
the charge and spin response. Figure \ref{fig2}(b) shows that their
respective spin changes are very different and always slower than the
charge response. The net reduction of the spin change is larger for
0.64 $\ub$ than 0.19 $\ub$. This is because once the spin moment is
small, there is no room for further reduction.

We can reveal some additional insights into the general trend through
a model.  For magnets, regardless of ferromagnets, antiferromagnets or
ferrimagnets, a minimum model must include the exchange
interaction. It is this interaction that sustains the long-range magnetic ordering across different lattice sites.  However,
including the exchange interaction is not enough to understand
magnetization changes since the total spin momentum is a conserved
quantity without spin-orbit coupling. We note in passing that in the
traditional magnetism theory, the spin moment change is built in from
the beginning through the Bose-Einstein distribution of magnons, so that
it does not apply here. With these considerations, we start
from the spin-orbit-coupled Heisenberg model \cite{jcp11}, \be
H_0=-\sum_{ij}J \hat{\vec{s}}_i\cdot \hat{\vec{s}}_j +\sum_i
\lambda \hat{\vec{l}}_i\cdot \hat{\vec{s}}_{i}
\label{ham}
\ee where $J$ is the exchange interaction
 between nearest-neighbor atomic sites $i$ and $j$,
$\lambda$ is the spin-orbit
coupling,
$\hat{\vec{s}}_i$ is the spin operator at
site $i$, and $\hat{\vec{l}}_i$ is the orbital operator at site $i$.
%\be
%H_I=\frac{e}{m}\sum_i \vec{p}_i\cdot \vec{A}(t)
%\ee
Within the Heisenberg picture, the $z$-component of the spin momentum
at site $i$ evolves according to \be \dot{\hat{s}}_{iz}=
\lambda(\hat{l}_{ix}\hat{s}_{iy}-\hat{l}_{iy}\hat{s}_{ix}) -\sum_j
J(\hat{s}_{iy}\hat{s}_{jx}-\hat{s}_{ix}\hat{s}_{jy}).
\label{spin}
\ee
The total spin momentum change $\dot{\hat{S}}_z$
is a sum over all the
sites, \be \dot{\hat{S}}_z=\sum_i\dot{\hat{s}}_{iz} = \sum_i
\lambda(\hat{l}_{ix}\hat{s}_{iy}-\hat{l}_{iy}\hat{s}_{ix}),
\label{spintotal}\ee where the
exchange interaction term drops out.
To reveal the role of the exchange coupling between
different sites, we first integrate similar equations like
Eq. (\ref{spin}) for $\hat{s}_{ix}$ and $\hat{s}_{iy}$, and then
substitute them back into Eq. (\ref{spintotal}).
  The resultant equation (\ref{spintotal}), that is
 linear in $\lambda$, is
\begin{widetext}
 \ba
\dot{\hat{S}}_z&\approx&\lambda \sum_i \left (\hat{l}_{ix}
\hat{s}_{iy}(-\infty)-\hat{l}_{iy} \hat{s}_{ix}(-\infty) \right) -
\lambda J \int^t_{-\infty}\sum_i \left
(\hat{l}_{ix}\hat{s}_{ix}+\hat{l}_{iy}\hat{s}_{iy} \right)\hat{S}_z
dt' \nonumber \\ &+&\lambda J
\int^t_{-\infty}\sum_i\hat{l}_{ix}\hat{s}_{iz}\hat{S}_x dt' +\lambda J
\int^t_{-\infty}\sum_i\hat{l}_{iy}\hat{s}_{iz}\hat{S}_y dt',
\label{spin2}
\ea
\end{widetext}
where $\hat{l}$ is a function of $t$, not $t'$, which should not be
integrated over. The first term on the right-hand side represents the
contribution from the orbital momentum. The second term is directly
proportional to the spin momentum $\hat{S}_z$ itself and represents
the spin relaxation. The last two terms are the precessional terms
since they are linked to the $x$- and $y$-components of the spin
momentum.  Next, we assume initially that the spin is along the
$z$-axis, so $\la \hat{s}_{ix}(-\infty)\ra=\la
\hat{s}_{iy}(-\infty)\ra=0$. To first order in the spin-orbit basis
\cite{prb09}, the last two terms are zero, so we also ignore them and
only keep terms that contain $\hat{S}_z$. This approximation is crude,
but in the beginning of spin dynamics, the demagnetization is
dominant, and the spin precession is expected to be small. Under these
approximations, we find \be \dot{\hat{S}}_z\approx -\frac{1}{2}\lambda
J \int^t_{-\infty} \sum_i \left
(\hat{l}^\dagger_i\hat{s}_i^-+\hat{l}_i^-\hat{s}_i^\dagger
\right)\hat{S}_z dt' \label{spin1} \label{master}.\ee This is the
master equation of the (de)magnetization process. For the first time,
the spin momentum change rate is linked to the spin $s$ and orbital
$l$ momenta, exchange interaction $J$ and spin-orbit coupling
$\lambda$.  The master equation shows that the spin momentum rate
depends on the history of the spin momentum itself, a non-Markovian
process. Note that both $s$ and $S$ affect the demagnetization rate.
We see that for a larger $S_z$, its (de)magnetization rate $\dot{S}_z$
is larger.  This means that for the same amount of spin moment change,
it needs less time, or a shorter demagnetization time. This is
consistent with our numerical calculation above. It is the direct
coupling of the spin to the orbital degree of freedom that allows the
spin to act upon itself self-consistently. The fact that the product
of spin-orbit coupling and exchange interaction enters the rate
equation highlights the critical role of the exchange interaction as
the chief protector for magnetic ordering and spin-orbit coupling as
the main channel for demagnetization. If $J$ is very small, we have
Eq. (\ref{spintotal}) to determine the demagnetization time, which is
normally very long, in particular in magnetic semiconductors. A strong
$J$ shortens the demagnetization time, which matches the
experimentally observed time scale. Because of this convoluted
interaction, the spin moment affects the demagnetization time
nonlinearly.

In conclusion, we have investigated how the demagnetization time
depends on the spin moment in ferromagnetic fcc Ni. We employ the
fixed spin moment method to systematically change the spin moment. For
each spin moment, we compute the laser-induced ultrafast spin moment
change as a function of time, from which we determine the
demagnetization time. Our results show that in general the
demagnetization time becomes shorter with a larger spin moment, and
the dependence is highly nonlinear. Quantitatively, we find that $\tm$
decreases with $M_z$ precipitously up to 0.35 $\ub$ before a small
increase around 0.45 $\ub$. A further decrease is observed
afterward. To reveal some further insights, we adopt the extended
Heisenberg model, which includes both the exchange interaction and
spin-orbit coupling. We demonstrate that the generic feature is indeed
reproduced. The reduction of the demagnetization time with the spin
moment originates from the sharp increase of the demagnetization rate
through the spin moment itself. The exchange interaction and
spin-orbit coupling jointly determine the time scale of the
demagnetization. The direct relation between the demagnetization time
and exchange interaction is consistent with the experimental
observation \cite{mathias}. A future experimental test of our
prediction is much needed, but is potentially challenging since other
parameters may contribute. However, one may dope the system with some
impurities \cite{mathias}, or employ the temperature \cite{ko}, laser
fluence \cite{koopmans}, or pressure \cite{tor}.  Since the
demagnetization time is an integrated part of femtomagnetism, we
believe that our finding will pave the way to reveal the intricate
mechanism of laser-induced ultrafast demagnetization.

\acknowledgments This work was solely supported by the U.S. Department
of Energy under Contract No.  DE-FG02-06ER46304.  Part of the work was
done on Indiana State University's Quantum cluster and High
performance computers.  This research used resources of the National
Energy Research Scientific Computing Center, which is supported by the
Office of Science of the U.S.  Department of Energy under Contract
No. DE-AC02-05CH11231. Our calculations also used resources of the
Argonne Leadership Computing Facility at Argonne National Laboratory,
which is supported by the Office of Science of the U.S. Department of
Energy under Contract No.  DE-AC02-06CH11357.

$^*$gpzhang@indstate.edu

\end{document}